\documentclass[aps,floatfix,showpacs,nofootinbib,showkeys,twocolumn,superscriptaddress]{revtex4-1}
\usepackage[T1]{fontenc}
\usepackage{amssymb,amsfonts,amsmath,mathtext,enumerate,float,mathrsfs}
\usepackage[utf8]{inputenc}
\usepackage[english]{babel}
\usepackage{graphicx,color}
\usepackage{wrapfig}
\graphicspath{{./}}

\usepackage{epsfig}
\usepackage{epsf}
\usepackage{subfig} 
\usepackage{feynmp} 
\usepackage{pgf,pgfarrows,pgfnodes,pgfautomata,pgfheaps,multirow}
\usepackage{array}
\usepackage{bm}
\usepackage{latexsym}
\usepackage{pifont}

\newcommand{\be}{\begin{equation}}
\newcommand{\ee}{\end{equation}}
\newcommand{\bea}{\begin{eqnarray}}
\newcommand{\eea}{\end{eqnarray}}
\newcommand{\dd}{\mbox{d}} 
\newcommand\nn{\nonumber}

\begin{document}


\title{The process $e^+e^- \to \eta(\eta')2\pi$ in the extended NJL model}

\author{M.\ K.\ Volkov}
\email{volkov@theor.jinr.ru}
\affiliation{Bogoliubov Laboratory of Theoretical Physics, JINR, Dubna, 141980 Russia}

\author{A.\ B.\ Arbuzov}
\email{arbuzov@theor.jinr.ru}
\affiliation{Bogoliubov Laboratory of Theoretical Physics, JINR, Dubna, 141980 Russia}
\affiliation{Department of Higher Mathematics, Dubna University, Dubna, 141980 Russia}

\author{D.\ G.\ Kostunin}
\email{dmitriy.kostunin@kit.edu}
\affiliation{Institut f\"ur Kernphysik, Karlsruhe Institute of Technology (KIT), Germany}

\begin{abstract}
Process of electron-positron annihilation into $\eta(\eta')2\pi$
is described within the extended NJL model in the energy range up to
about 2~GeV. Contributions of intermediate vector mesons $\rho(770)$
and $\rho(1450)$ are taken into account.
Results for the $\eta2\pi$ channel are found to be in a reasonable agreement with
experimental data. Predictions for production of $\eta' 2\pi$ are given. 
The corresponding estimations for decays $\tau\to\eta(\eta')2\pi\nu$ are given in Appendix.
\end{abstract}

\date{\today}

\keywords{electron-positron annihilation, Nambu-Jona-Lasinio model}

\pacs{
13.60.Le,
13.66.Bc,
12.39.Fe,
}

\maketitle

\section{Introduction}

The description of hadronic interactions at low energies is problematic.
Indeed, perturbative QCD hardly works at energies below 2~GeV. 
So in this domain various phenomenological models are used, most of them
are based on the chiral symmetry of strong interactions. One of the most popular models
of such a kind is the Nambu--Jona-Lasinio (NJL)
one~\cite{Ebert:1982pk,Volkov:1984kq,Volkov:1986zb,Ebert:1985kz,Vogl:1991qt,Klevansky:1992qe,Volkov:1993jw,Ebert:1994mf}.
This model describes spectra of light mesons in the ground states
and their interactions using a rather small number of parameters.

Recently in the framework of the extended NJL model, a series of processes
of meson production in electron-positron annihilation was
described~\cite{Arbuzov:2010xi,Arbuzov:2011zz,Ahmadov:2011ey,Volkov:2012tk,Ahmadov:2013ksa}.
In the corresponding calculations, we took into account contributions
of intermediate vector mesons $\rho(770)$, $\omega(782)$, $\phi(1020)$
and the radial excited states $\rho(1450)$, $\omega(1420)$.
The radial excited states are treated with the help of the extended
NJL model suggested in papers~\cite{Volkov:1996br,Volkov:1996fk,Volkov:1997dd,Volkov:1999yi,Volkov:2006vq}.
It was demonstrated that the extended NJL model provides a reasonably good
description of a wide class of strong interaction processes at energies up to
about 1.5~GeV.

In the present paper we finalize a series of studies by consideration
of the reaction $e^+e^-\to\eta(\eta')2\pi$. Both $\rho(770)$ and $\rho(1450)$
intermediate states are taken into account. The applicability of our
calculation is limited to the domain of the center-of-mass energies up to
about 2~GeV.

The process $e^+e^-\to\eta2\pi$ has been studied experimentally at several
facilities: DM1~\cite{Cordier:1979qg}, DM2~\cite{Antonelli:1988fw},
ND~\cite{Druzhinin:1986dk,Dolinsky:1991vq}, CMD-2~\cite{Akhmetshin:2000wv},
and BaBar~\cite{Aubert:2007ef}.
From the theoretical point of view they were also discussed within several
different phenomenological approaches~\cite{Akhmetshin:2000wv,Dumm:2012vb,Dai:2013joa}.
Comparison of our results with the ones presented in those papers will be given
in Conclusions.
Below we present the corresponding description in the framework of the
extended NJL model and give a comparison with experimental data.

\section{The extended NJL model}

The Lagrangian of quark-meson interactions in the extended NJL model was given
in Refs.~\cite{Volkov:1996fk,Volkov:1997dd,Volkov:2006vq,Volkov:2012tk,Ahmadov:2013ksa}.
After bosonization
and diagonalization of the free field Lagrangian, the relevant quark-meson interactions take
the form\footnote{Let note that for $\eta$ and $\eta'$ mesons we will use only part of Lagrangian contains interactions with $u$ and $d$ quarks,
the full Lagrangian can be found in~\cite{Volkov:2006vq,Volkov:1999xf}.}
\begin{eqnarray}
\Delta {\mathcal L}_2^{\mathrm{int}} &=& \bar{q}(k')\biggl( L_{\mathrm{f}} + L_\gamma + L_\mathrm{V} + L_{\pi, \hat\pi} + L_{\boldsymbol\eta} \biggr) q(k),\\
L_{\mathrm{f}} &=& i\hat\partial - m, \nn \\
L_\gamma &=& \frac{e}{2}\left(\tau_3+\frac{\mathrm{I}}{3}\right)\hat{A}, \nn \\
L_\mathrm{V} &=& A_{\rho}\tau_3{\hat{\rho}}(p) - A_{\rho'}\tau_3{\hat\rho'}(p), \nn \\
L_{\pi,\pi'} &=& A_\pi \tau_{\pm}\gamma_5\pi(p) - A_{\pi'} \gamma_5\tau_{\pm}\pi'(p), \nn \\
L_{\boldsymbol\eta} &=& i\gamma_5\mathrm{I}\sum\limits_{\boldsymbol\eta = \eta,\eta',\hat\eta,\hat\eta'}A_{\boldsymbol{\eta}}\boldsymbol\eta(p), \nn
\end{eqnarray}
where $\bar{q}=(\bar{u},\bar{d})$ with $u$ and $d$ quark fields;
$m=\mathrm{diag}(m_u,m_d)$, $m_u=m_d=280$~MeV are the constituent quark masses;
$e$ is the electron charge;
$\hat A$ is the photon field; $\rho$, $\omega$ ($\rho'$, $\omega'$), $\pi$ ($\pi'$), $\eta$, $\eta'$ ($\hat\eta$, $\hat\eta'$) are meson fields (hats over $\eta$ and $\eta'$ mean exited states);
$\tau^{\pm} = (\tau_1 \mp i\tau_2)/\sqrt{2}$, $\tau_{1,2,3}$ are Pauli matrices; and $\mathrm{I}$ is the unit matrix.
Quantities $A_i$ read
\begin{eqnarray}
A_{\rho} &=& g_{\rho_1}\frac{\sin(\beta+\beta_0)}{\sin(2\beta_0)}
       +g_{\rho_2}f({k^\bot}^2)\frac{\sin(\beta-\beta_0)}{\sin(2\beta_0)},
\\
A_{\rho'} &=& g_{\rho_1}\frac{\cos(\beta+\beta_0)}{\sin(2\beta_0)}
        +g_{\rho_2}f({k^\bot}^2)\frac{\cos(\beta-\beta_0)}{\sin(2\beta_0)},
\nn \\
A_{\pi} &=& g_{\pi_1}\frac{\sin(\alpha+\alpha_0)}{\sin(2\alpha_0)}
       +g_{\pi_2}f({k^\bot}^2)\frac{\sin(\alpha-\alpha_0)}{\sin(2\alpha_0)},
\nn \\
A_{\pi'} &=& g_{\pi_1}\frac{\cos(\alpha+\alpha_0)}{\sin(2\alpha_0)}
        +g_{\pi_2}f({k^\bot}^2)\frac{\cos(\alpha-\alpha_0)}{\sin(2\alpha_0)}, \nn \\
A_{\boldsymbol \eta} &=& g_{\pi_1}\varphi_{\boldsymbol \eta}^{1} + g_{\pi_2}\varphi_{\boldsymbol \eta}^{2}f({k^\bot}^2). \nn
\end{eqnarray}
Radially-excited states are described in the extended NJL model using the following form factor in the quark-meson interaction:
\begin{eqnarray}
&& f({k^\bot}^2) = (1-d |{k^\bot}^2|) \Theta(\Lambda_3^2-|{k^\bot}^2|),
\\
&& {k^\bot} = k - \frac{(kp)p}{p^2},\,\,\ d = 1.788\ {\mathrm{GeV}}^{-2}, \nn
\end{eqnarray}
where $k$ and $p$ are the quark and meson momenta, respectively;
$\Lambda_3=1.03$~GeV is the cut-off parameter.
The coupling constants are defined in the extended NJL model by the integrals containing given form-factors
\begin{eqnarray}
&& g_{\pi_1} = \left(4 \frac{I_2^{(0)}}{Z}\right)^{-1/2} = 3.01, \\
&& g_{\pi_2} = \left(4 I_2^{(2)}\right)^{-1/2} = 4.03, \nn \\
&& g_{\rho_1} = \left(\frac{2}{3} I_2^{(0)}\right)^{-1/2} = 6.14,\nn \\ 
&& g_{\rho_2} = \left(\frac{2}{3} I_2^{(2)}\right)^{-1/2} = 9.87, \nn
\end{eqnarray}
where $Z$ factor appeared after taking into account pseudoscalar -- axial-vector transitions, $Z \approx 1.2$.
Note that $g_{\pi_1} \approx m_u/F_\pi$, where $F_\pi \approx 93$ MeV is the pion decay constant.
The quark loop integrals are defined as:
\begin{equation}
I^{(n)}_m = -iN_c\int\frac{\dd^4 k}{(2\pi)^4}\frac{(f({k^\bot}^2))^n}{(m_u^2-k^2)^m}\Theta(\Lambda^2_3 - \vec k^2),
\end{equation}
where $N_c = 3$ is the number of colors. The mixing angles for pseudoscalar and vector mesons are: $\alpha_0 = 58.39^{\circ}$, $\alpha = 58.70^\circ$, $\beta_0 = 61.44^{\circ}$, $\beta = 79.85^{\circ}$. One can find the definition of mixing angles for $\pi$ and $\rho$ mesons
in~\cite{Volkov:1996fk,Volkov:1997dd}. The mixing coefficients for the isoscalar pseudoscalar meson states given in Table~\ref{table:1}
were derived in~\cite{Volkov:1999xf,Volkov:1999yi,Volkov:2006vq}.
\begin{table}
$$
	\begin{array}{|r|c|c|c|c|}
	\hline
	 		&\eta 		&\hat\eta 	&\eta' 		&\hat\eta'\\
	\hline
	\varphi_{\boldsymbol{\eta}}^{1}	&0.71		&0.62		&-0.32		&0.56		\\
	\varphi_{\boldsymbol{\eta}}^{2}   &0.11		&-0.87		&-0.48		&-0.54		\\
	\hline
	\end{array}
$$
\caption{Mixing coefficients for isoscalar pseudoscalar meson states ($\boldsymbol{\eta} = \eta, \eta', \hat \eta, \hat \eta'$).}
\label{table:1}
\end{table}

For the simplification of the presentation we define:
\begin{eqnarray}
&& \varphi_\pi = \frac{1}{\sin(2\alpha_0)}\left(\begin{array}{c}\sin(\alpha + \alpha_0)\\ \sin(\alpha - \alpha_0)\end{array}\right),\\
&& \varphi_\eta = \left(\begin{array}{c}0.71\\0.11\end{array}\right),\quad \varphi_{\eta'} = \left(\begin{array}{c}-0.32\\-0.48\end{array}\right), \nn \\
&& \varphi_\rho = \frac{1}{\sin(2\beta_0)}\left(\begin{array}{c}\sin(\beta + \beta_0)\\ \sin(\beta - \beta_0)\end{array}\right), \nn \\
&& \varphi_{\rho'} = -\frac{1}{\sin(2\beta_0)}\left(\begin{array}{c}\cos(\beta+\beta_0)\\ \cos(\beta - \beta_0)\end{array}\right). \nn
\end{eqnarray}

\section{Process Amplitudes and Cross Section}

\begin{figure}[h]
\begin{center}
\includegraphics[width=0.7\linewidth]{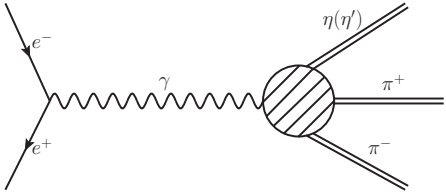}
\end{center}
\caption{The Feynman diagram with an intermediate photon. The dashed circle represents the sum of two
sub-diagrams given in Figs.~\ref{diag3}~and~\ref{diag4}.}
\label{diag1}
\end{figure}

\begin{figure}[h]
\begin{center}
\includegraphics[width=0.7\linewidth]{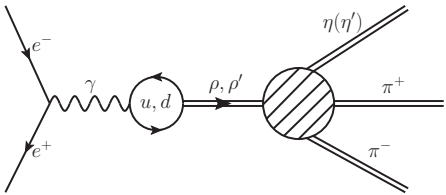}
\end{center}
\caption{The Feynman diagram with intermediate $\rho(770)$ and $\rho(1450)$ vector mesons. The dashed circle represents the sum of two
sub-diagrams given in Figs.~\ref{diag3}~and~\ref{diag4}.}
\label{diag2}
\end{figure}

The total amplitude of the given process has the form
\begin{equation}
\label{total_amp}
T = -\frac{4\pi\alpha}{q^2}\bar e \gamma^\mu e \mathcal{H}_\mu\,,
\end{equation}
where $q = p_{e^{+}} + p_{e^{-}}$ in the center-of-mass system.
The hadronic part of the amplitude takes the form\footnote{Hereafter $\boldsymbol{\eta} = \eta, \eta'$.}
\begin{eqnarray}
\label{hadr_amp1}
&& \mathcal{H}_\mu = V_\mu\left(T_\gamma(q^2,s) + \sum\limits_{V=\rho,\rho'}T_V(q^2,s)\right), \\
&& V_\mu = p^\alpha_{\boldsymbol\eta} p^\beta_{\pi^+} p^\gamma_{\pi^-}\varepsilon_{\mu\alpha\beta\gamma}. \nn
\end{eqnarray}
Electron-positron annihilation with $\boldsymbol{\eta}\pi\pi$ production is described by Feynman diagrams
with virtual photons, Fig.~\ref{diag1}, and with intermediate vector $\rho(770)$ and $\rho(1450)$ mesons, Fig.~\ref{diag2}.
In the following calculations we took into account the ground state $\rho(770)$ and the first radial-exited state $\rho(1450)$
\begin{eqnarray}
\label{hadr_amp2}
T_\gamma(q^2,s) &=& \sum\limits_{i=1}^2 g_{\pi_i}\varphi_{\boldsymbol\eta}^i \left(T^{(i-1)}_\Box(s) + T^{(i-1)}_\triangle(s)\right)\,, \\
T_V(q^2,s) &=& \frac{(C_{\gamma V} / g_{V_1})q^2}{m^2_V - q^2 - i\sqrt{q^2}\Gamma_V(q^2)} \nn \\ 
&\times& \sum\limits_{i=1}^2\sum\limits_{j=1}^2 g_{\pi_i}\varphi_{\boldsymbol\eta}^i g_{V_j}\varphi_{V}^j \left(T^{(i+j-2)}_\Box(s) + T^{(i+j-2)}_\triangle(s)\right). \nn
\end{eqnarray}

As was shown in previous calculations (see for example~\cite{Volkov:2012tk,Ahmadov:2013ksa}), vector meson dominance could be directly obtained in the standard NJL model. In present calculations we use only the extended version of the model, which describes the $\gamma V$ transition only with an accuracy about 10\%. Thus, for the $\rho(770)$ resonance we apply vector meson dominance directly:
\begin{equation}
T_\gamma(q^2,s) + T_\rho(q^2,s) \to T_\rho^{\mathrm{VMD}}(q^2,s) = \frac{m^2}{q^2} T_\rho(q^2,s).
\end{equation}

\begin{figure}[h]
\begin{center}
\includegraphics[width=0.45\linewidth]{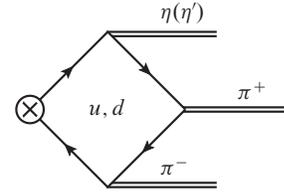}
\end{center}
\caption{The $V\eta\pi\pi$ vertex with quark loop of the box type. Interchange of pseudoscalar meson lines gives factor 3!.}
\label{diag3}
\end{figure}

\begin{figure}[h]
\begin{center}
\includegraphics[width=0.55\linewidth]{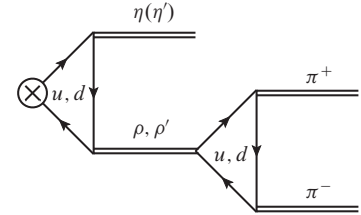}
\end{center}
\caption{The $V\eta\pi\pi$ vertex with two triangle quark loops connected by a virtual vector meson.}
\label{diag4}
\end{figure}

\begin{widetext}
The vertices $\gamma{\boldsymbol\eta}\pi\pi$ and $V{\boldsymbol\eta}\pi\pi$ contain the sum of two terms
\begin{eqnarray}
\label{vertex}
T^{(n)}_\Box(s) &=& -24 F_\pi g_\pi^3 I^{(n)}_4, \\
T^{(n)}_\triangle(s) &=& 16 F_\pi g_\pi \sum\limits_{V = \rho,\rho'} \frac{g_{V\to\pi\pi}}{m^2_V - s - i\sqrt{s}\Gamma_V(s)} \sum\limits_{i=1}^2  \nn g_{\rho_i}\varphi_{V}^i I_3^{(n+i-1)} \approx 16
F_\pi g_\pi \frac{g_{\rho\to\pi\pi}}{m^2_\rho - s - i\sqrt{s}\Gamma_\rho(s)} \sum\limits_{i=1}^2  g_{\rho_i}\varphi_{\rho}^i I_3^{(n+i-1)}. \nn
\end{eqnarray}
\end{widetext}
The term $T^{(n)}_\Box(s)$ corresponds to the contribution of quark box-diagram, see Fig.~\ref{diag3}. The term $T^{(n)}_\triangle(s)$
comes from two triangle quark loops connected by a virtual vector meson, see Fig.~\ref{diag4}. We neglected the contribution 
of the $\rho(1450)$ in the $T^{(n)}_\triangle(s)$ term, since it is very much suppressed with respect to the $\rho(770)$ contribution by kinematics and also
due to the small partial decay width of $\rho(1450)\to2\pi$ see~\cite{Volkov:1997dd}.

Since $g_{\pi_1}\varphi^1_\pi \gg g_{\pi_2}\varphi^2_\pi \approx 0$, we replaced all terms contain vertices with $\pi$ mesons
(see discussion in Refs.~\cite{Volkov:1997dd,Volkov:2012tk})
\begin{equation}
\prod\limits_{i=1}^n\sum\limits_{j=1}^2 g_{\pi_j} \varphi_\pi^j T_{\mathrm{non-}\pi}^{(k)} I_{n+k}^{(k+ij-i)} \biggr|_{g_{\pi_2}\varphi^2_\pi\to 0} =
g_{\pi_1}^nT_{\mathrm{non-}\pi}^{(k)} I_{n+k}^{(k)}.
\label{pi_simpl}
\end{equation}
The second triangle diagram describing decay $V\to\pi\pi$ was computed in the framework of the extended NJL model in Refs.~\cite{Volkov:1997dd,Volkov:2012tk}, it gives:
\begin{equation}
g_{V\to\pi\pi} \approx g_{\rho_1}\varphi_{V}^1 + g_{\rho_2}\varphi_{V}^2\frac{I_2^{(1)}}{I_2^{(0)}}.
\end{equation}



The transitions of a photon into the vector mesons ($\rho$,$\rho'$) denoted by the terms
\begin{equation}
C_{\gamma V} = \varphi_{V}^1 + \varphi_{V}^2\frac{I_2^{(1)}}{\sqrt{I_2^{(0)} I_2^{(2)}}}.
\end{equation}
\begin{widetext}
We chose the fixed width for $\rho(770)$ and the running one~\cite{Arbuzov:2011zz,Volkov:2012tk} for $\rho(1450)$:
\begin{eqnarray}
\Gamma_\rho(s) &=& \Gamma_\rho, \\
\Gamma_{\rho'}(s) &=& \Theta(2m_\pi - \sqrt{s}) \Gamma_{\rho'\to2\pi} \nn
\\ &+& \Theta(\sqrt{s} - 2m_\pi)\biggl(\Gamma_{\rho'\to 2\pi} + \Gamma_{\rho'\to\omega\pi}\frac{\sqrt{s}-2m_\pi}{m_\omega-m_\pi}\biggr)\Theta(m_\omega + m_\pi - \sqrt{s})
\nn
\\ &+& \Theta(m_{\rho'} - \sqrt{s})\Theta(\sqrt{s} - m_\omega - m_\pi)
 \left(\Gamma_{\rho'\to 2\pi} + \Gamma_{\rho'\to\omega\pi}
 + (\Gamma_{\rho'} - \Gamma_{\rho'\to2\pi}  -\Gamma_{\rho'\to\omega\pi})
\nn
\frac{\sqrt{s}- m_\omega - m_\pi}{m_{\rho'} - m_\omega - m_\pi}\right)
\nn
\\ &+& \Theta(\sqrt{s} - m_{\rho'})\Gamma_{\rho'}(m^2_{\rho'}), \nn
\end{eqnarray}
\end{widetext}
where $\Gamma_\rho = 147.8$~MeV and $\Gamma_{\rho'}(m^2_{\rho'}) = 400$~MeV are taken from PDG~\cite{Beringer:1900zz}.
The values $\Gamma(\rho'\to2\pi)=22$~MeV and $\Gamma(\rho'\to\omega\pi^0)=75$~MeV were calculated in \cite{Volkov:1997dd}.

\begin{figure}[t]
\begin{center}
\includegraphics[width=1.0\linewidth]{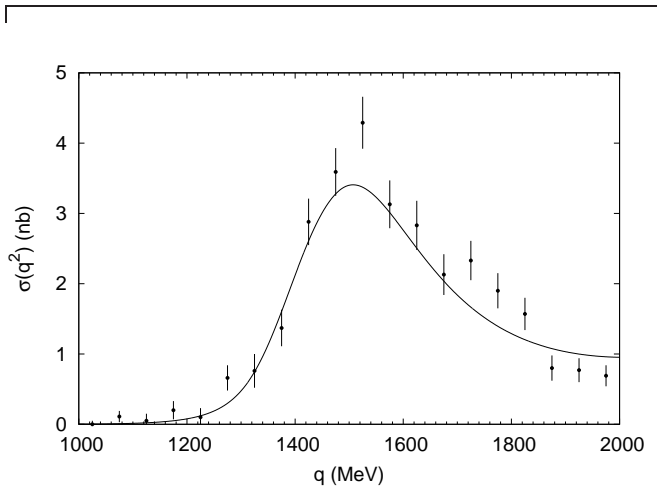}
\end{center}
\caption{Comparison of the extended NJL model predictions with the BaBar experiment data~\cite{Aubert:2007ef} for $e^+e^-\to\eta 2\pi$ process.}
\label{cs_eta}
\end{figure}

\begin{figure}[t]
\begin{center}
\includegraphics[width=1.0\linewidth]{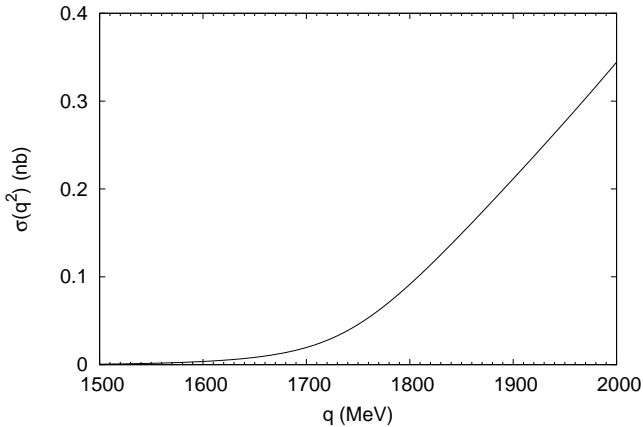}
\end{center}
\caption{Prediction of the extended NJL model for $e^+e^-\to\eta' 2\pi$ process.}
\label{cs_eta1}
\end{figure}

The total cross-section takes the form
\begin{equation}
\label{sigma}
\sigma(q^2) = \frac{\alpha^2}{192\pi q^6}\int\limits^{s_+}_{s_-} \dd s\int\limits^{t_+}_{t_-} \dd t |T(q,s,t)|^2\,,
\end{equation}
where variables are defined as $s = (p_{\boldsymbol\eta} + p_{\pi^{+}})^2$, $t = (p_{\boldsymbol\eta} + p_{\pi^{-}})^2$, and the limits are
\begin{eqnarray}
&& t_{\mp} = \frac{1}{4s}\biggl( [q^2 + m_{\boldsymbol\eta}^2 - 2m_\pi^2]^2 - \nn \\ && \qquad - [\lambda^{1/2}(q^2,s,m_\pi^2) \pm \lambda^{1/2}(m_{\boldsymbol\eta}^2, m_\pi^2, s)]^2 \biggr)\,, \\
&& s_{-} = (m_{\boldsymbol\eta} + m_\pi)^2, \qquad
s_{+} = (\sqrt{q^2} - m_\pi)^2, \nn \\
&& \lambda(a,b,c) = (a - b - c)^2 - 4bc. \nn
\end{eqnarray}
The masses of all particles were taken from PDG~\cite{Beringer:1900zz}: $m_{\pi^{\pm}} = 139.57$~MeV, $m_\eta = 547.86$~MeV, $m_{\eta'} = 957.78$~MeV, $m_\rho = 775.49$~MeV, $m_{\rho'} = 1465$~MeV. One can see the final results in Figs.~\ref{cs_eta},~\ref{cs_eta1}.

\section{Discussion and Conclusions}

The presented calculation shows that the extended NJL model allows
to describe the energy dependence of the total cross section of $e^+e^-$
annihilation into $\eta 2\pi$ in a satisfactory agreement with experimental data
in the energy region up to 2~GeV. This allows to expect that our predictions for the
channel $e^+e^-\to\eta' 2\pi$ are also reasonable.

One of the first attempts to provide a theoretical interpretation of the
experimental data for $e^+e^-\to\eta 2\pi$ was presented in paper~\cite{Akhmetshin:2000wv}. 
The vector meson dominance model was used taking into account intermediate vector mesons $\rho(770)$,
$\rho(1450)$, and $\rho(1700)$. A number of free parameters was fitted from the experimental
data on the same process. Only one structure which corresponds in our case to the
$T_\triangle(s)$ in the amplitude of the process was considered.

In paper~\cite{Dumm:2012vb} a resonance chiral theory was used. This model contains a very large
number of free parameters. However, the contributions of intermediate radial-excited $\rho$ mesons
were not taken into account there, 
while the one due to $\rho(1450)$ meson is certainly important in the description of this process in the energy
region under consideration.
An advanced application of the same model was presented in Ref.~\cite{Dai:2013joa},
where intermediate vector mesons $\rho(770)$, $\rho(1450)$, and $\rho(1700)$
were included by means of introduction of additional free parameters. Note that
the contribution of $\rho(1700)$ was found to be not very important numerically.

The main difference of our results from the previous ones is that we work in the
extended NJL model where all the parameters have been fixed from the beginning.
Moreover, for the width of $\rho(1450)$ we take the PDG value 400~MeV, while in 
the alternative approaches considerably lower values were used, namely:
211~MeV in~\cite{Akhmetshin:2000wv} and 238~MeV in~\cite{Dai:2013joa}.  
As the result our model has a certain predictive power, contrary to the alternative approaches.
We would like to underline that the extended NJL model was extensively tested by description of a large
series of different processes with strong, weak, and electromagnetic interactions of
mesons~\cite{Volkov:1999yi,Arbuzov:2011zz,Ahmadov:2013ksa,Volkov:2012tk,
Arbuzov:2010xi,Ahmadov:2011ey,Volkov:2001ct,Volkov:2012uh,Volkov:2012gv,Volkov:2012be}.

\section*{Appendix. The decays $\tau\to\eta(\eta')2\pi\nu$}

To proceed the calculation of decay $\tau\to\eta(\eta')2\pi\nu$ we complement the Lagrangian 
with terms contain $W$-boson and charged $\rho$ and $\rho'$ mesons
\begin{eqnarray}
\Delta {\mathcal L}_2 &=& \frac{g_{EW}}{\sqrt 2}\hat{W} + A_{\rho}\tau_{\pm}{\hat{\rho}}(p) 
- A_{\rho'}\tau_{\pm}{\hat\rho'}(p),
\end{eqnarray}
where $\tau_{\pm} = (\tau_1 \mp i\tau_2)/\sqrt{2}$, and $g_{EW}$ is electoweak coupling constant.

\begin{table}[h!]
\caption{Branching ratios for the processes $\tau\to\eta(\eta')2\pi\nu$}
\begin{center}
\begin{ruledtabular}
\begin{tabular}{cccc}
Process & Full amplitude & Only $\rho(770)$ & PDG~\cite{Beringer:1900zz} \\
 $ \mathcal B(\tau\to\eta2\pi\nu)\cdot 10^3 $ & $1.46$ & $1.01$ & $1.39 \pm 0.10$ \\
 $ \mathcal B(\tau\to\eta'2\pi\nu)\cdot 10^5 $ & $ 0.09 $ & $0.12$ & $ < 1.2$ \\
\end{tabular}
\end{ruledtabular}
\end{center}
\label{tbl_comp}
\end{table}

To get the amplidute for the present process we replaced the $e^+e^-$ current by $\tau\nu$ 
and intermediate $\gamma$ by $W^+$ gauge boson in the amplitude~(\ref{total_amp}) (Figs.~\ref{diag1} 
and \ref{diag2}). One can obtain the expression for the $\tau$ decay from electron-positron 
annihilation after applying the corresponding phase-space volume transformation
\begin{eqnarray}
\Gamma(\tau\to\eta(\eta')2\pi\nu) &=& \frac{3|V_{ud}|^2}{2\pi\alpha^2m_\tau^8}
\Gamma(\tau\to e\nu_e\nu_\tau)\cdot \\ && \int^{m_\tau}_0 \sigma(q^2)  
q^2(m_\tau^2-q^2)^2(m_\tau^2+2q^2) \dd q^2\,, \nn
\end{eqnarray}
where $\sigma(q)$ is the same as in~(\ref{sigma}), and $\Gamma(\tau\to e\nu_e\nu_\tau)$ takes the form
\begin{equation}
\Gamma(\tau\to e\bar\nu_e\nu_\tau) = \frac{G_f^2 m_\tau^5}{192\pi^3}.
\end{equation}
To compare with theoretical predictions which take into account only $\rho(770)$ 
resonance~\cite{Dumm:2012vb,braaten} we also give a prediction with $T_{\rho'} = 0$. 
Our estimations are in good agreement with current experimental data~\cite{Beringer:1900zz} 
(see Table~\ref{tbl_comp}).

\section*{Acknowledgments}
We are grateful to E.A.~Kuraev and N.M.~Plakida for useful discussions.

\end{document}